\documentclass[twoside]{article}
\usepackage{fleqn,espcrc2}
\usepackage{epsf}

\def\lsim{\raise0.3ex\hbox{$<$\kern-0.75em\raise-1.1ex\hbox{$\sim$}}}
\def\gsim{\raise0.3ex\hbox{$>$\kern-0.75em\raise-1.1ex\hbox{$\sim$}}}

\newcommand{\pslash}{p\kern-1ex /}
\newcommand{\Dslash}{{\cal D}\kern-1.5ex /}

\newcommand{\J}[4]{{#1} {\bf #2} (#3) #4}

\newcommand{\NP}{Nucl.~Phys.}
\newcommand{\NPSup}{Nucl.~Phys.~B (Proc.~Suppl.)}

\newcommand{\PR}{Phys.~Rev.}
\newcommand{\PRL}{Phys.~Rev.~Lett.}

\newcommand{\comment}[1]{}

\title{
\vspace*{-1.9cm}
\begin{flushright}
{\normalsize UTHEP-480}\\
{\normalsize UTCCP-P-144}\\
\end{flushright}
Non-perturbative renormalization of vector and axial vector currents
in quenched QCD for a renormalization group improved gauge action
\thanks{Talk presented by K.~Ide.}}

\author{
CP-PACS Collaboration: 
K.~Ide\rlap,$^{\rm a}$
S.~Aoki\rlap,%
\address{Institute of Physics, University of Tsukuba,
        Tsukuba, Ibaraki 305-8571, Japan}
M.~Fukugita\rlap,%
\address{Institute for Cosmic Ray Research, University of Tokyo,
        Kashiwa, Chiba 277-8582, Japan}
N.~Ishizuka\rlap,$^{\rm a,}$%
\address{Center for Computational Physics, University of Tsukuba,
        Tsukuba, Ibaraki 305-8577, Japan}
Y.~Iwasaki\rlap,$^{\rm a,c}$
K.~Kanaya\rlap,$^{\rm a}$
T.~Kaneko\rlap,%
\address{High Energy Accelerator Research Organization (KEK),
        Tsukuba, Ibaraki 305-0801, Japan}
Y.~Kuramashi\rlap,$^{\rm d}$
V.~Lesk\rlap,$^{\rm c}$
M.~Okawa\rlap,%
\address{Department of Physics, Hiroshima University,
        Higashi-Hiroshima, Hiroshima 739-8526, Japan}
Y.~Taniguchi\rlap,$^{\rm a}$
A.~Ukawa$^{\rm a,c}$ and
T.~Yoshi\'e$^{\rm a,c}$
}

\begin{document}

\pagestyle{empty}


\begin{abstract}

Renormalization constants of vector ($Z_V$) and axial-vector ($Z_A$)
currents are determined non-perturbatively in quenched QCD for an
RG-improved gauge action and a tadpole-improved
clover quark action using the Schr\"odinger functional method.
Meson decay constants $f_\rho$ and $f_\pi$ show much better scaling 
when $Z_V$ and $Z_A$ estimated for infinite physical volume are used 
instead of $Z$-factors from tadpole-improved one-loop perturbation theory.

\end{abstract}

\maketitle

\section{Introduction}

In a recent comprehensive study by the CP-PACS Collaboration 
of $N_f$=2 full QCD~\cite{ref:CPPACS-NF2}, 
meson decay constants were found to exhibit a very large scaling 
violation over the range of lattice spacing $a^{-1}\approx 1-2$GeV.
This was disappointing since an RG-improved gluon action and 
Sheikoleslami-Wohlert quark action with tadpole-improved $c_{\rm sw}$
were used.
In this calculation, however, one-loop perturbative $Z$-factors, 
albeit tadpole-improved, were used for currents. A natural question was 
whether scaling becomes improved if non-perturbative  
$Z$-factors are employed instead. 

At Lattice2001, we reported an initial study of this problem 
using the Schr\"odinger functional (SF) method~\cite{ref:CPPACS-NPZ}   
within quenched QCD. We found the problem of anomalously large values 
appearing in the ensemble of hadron correlators toward strong 
coupling  where CP-PACS data of decay constants had been taken. 
In this report, we have analyzed this problem in some detail. 
Here we present our final results on the $Z$-factors including 
these analysis.

\section{Method}\label{sec:calculation}

We follow the method developed by the ALPHA collaboration~\cite{ref:ALPHA-Z}, 
and work with a lattice geometry of $L^3\times T$ with $T=2L$
for $Z_V$ with a vector operator at $t=L$, and
for $Z_A$ with two axial vector operators at
$t=3T/8$ and $t=5T/8$.

Tree-level values are used for the coefficients of boundary counter
terms of the action.
For improving the axial current, we adopt the one-loop perturbative value
for the coefficient $c_A$.

Values of $Z_V$ and $Z_A$ are determined for $\beta=2.2$ -- 8.0
which approximately covers the range of the CP-PACS quenched 
calculation~\cite{ref:CPPACS-NF2}, $\beta =2.187$ -- 2.575. 
We have analyzed 200--20000 configurations depending on $\beta$
value and lattice size. 

\section{Exceptional Configurations}\label{sec:exceptional}
\begin{figure}[t]
\unitlength=1cm
\centerline{\epsfxsize=6.4cm \epsfbox{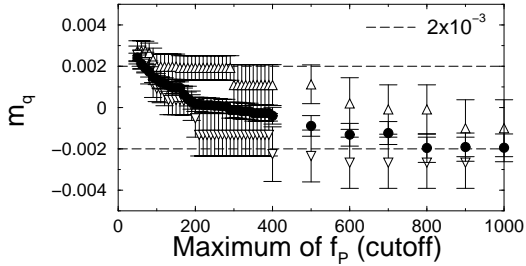}}
\vspace{-1.1cm}
\caption{Cutoff dependence of $m_q$ at $\beta=2.4$ on 
an $8^3\times 16$ lattice for three $\kappa$'s around $\kappa_c$.
 }
\label{fig:m-dep}
\vspace{-0.7cm}
\end{figure}

It is straight-forward to calculate $Z$-factors for $\beta \ge 2.6$.
For lower $\beta$ values on large lattices such as $8^3\times 16$, however,  
anomalously large values appear in the ensemble of hadron correlators.  
This makes it difficult to determine quark mass precisely, 
and since this means uncertainties in $\kappa_c$, also that of $Z$-factors.  

We suspect that these ``exceptional'' configurations are an artifact 
of quenched approximation; having very small or negative eigenvalues of 
the Wilson-Dirac operator, they would be suppressed in full QCD. 
Since one cannot distinguish ``exceptional'' configurations from 
``normal'' ones on some rigorous basis, we restrict the configurations 
used for averaging to those having the 
value of a relevant hadron correlator below some cutoff. 
We then examine if uncertainties under variation of the cutoff are contained 
within some acceptable magnitude. 

In Fig.~\ref{fig:m-dep} we illustrate this test for $m_q$ for which 
a cutoff is set for $f_P$ (see Ref.~\cite{ref:ALPHA-Z} for definition).  
We estimate $\kappa_c$ from $m_q$ with the cutoff value of 300,
because $m_q$ is rather stable there.
The uncertainty in $m_q$ at the $\kappa_c$ is $\approx \pm 2\times 10^{-3}$, 
once the cutoff of $f_P$ is taken in the range 200 -- 1000.

\begin{figure}[t]
\unitlength=1cm
\begin{picture}(6.4,5)(0,0)
\centerline{\epsfxsize=6.4cm \epsfbox{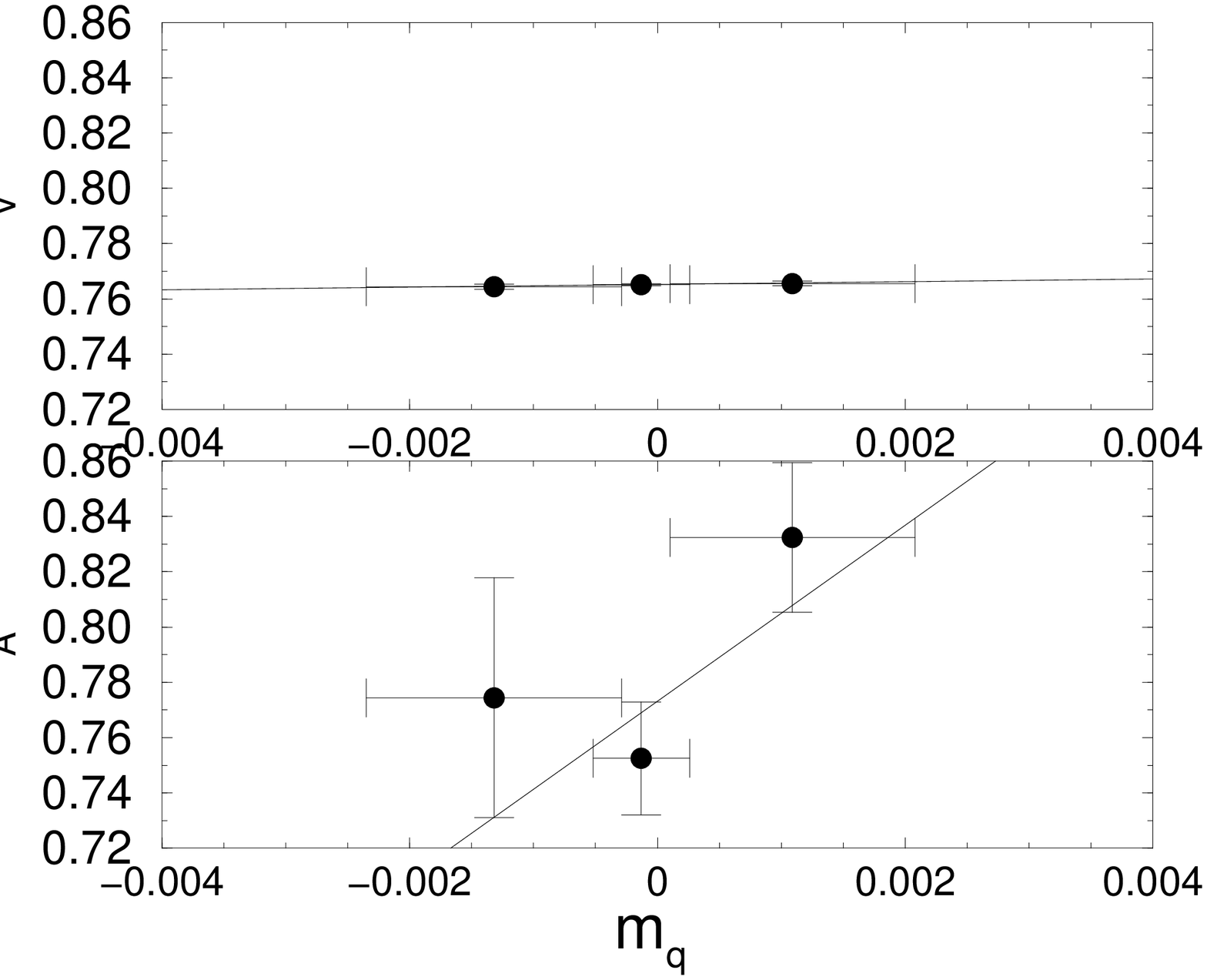}}
\put(-5.0,4.4){
\vector(-1,0){0}
\line(1,0){2.64}
\vector(1,0){0}
}
\put(-5.1,4.5){cut=200-1000 region}
\end{picture}
\vspace{-0.7cm}
\caption{$m_q$ dependence of $Z_V$ and $Z_A$ at $\beta=2.4$ on an $8^3\times 16$ lattice.}
\label{fig:mZVZA}
\vspace{-0.7cm}
\end{figure}

In Fig.~\ref{fig:mZVZA} we show how much the $Z$-factors depend on $m_q$. 
$Z_V$ is insensitive to $m_q$, and $Z_A$ is consistent within 10\% or so, 
albeit apparently exhibiting a more pronounced dependence.  

We analyze the uncertainties in the statistical averaging of $Z$-factors 
themselves by applying a cutoff in $f_1$, as carried out 
in Ref.~\cite{ref:CPPACS-NPZ}.
The conclusion is similar; $Z_V$ is very stable against variation of 
the cutoff, and $Z_A$ shows a more conspicuous variation of 5\% or so. 

Uncertainties of $Z_A$ on an $8^3$ lattice of order 15\% in total 
lead to uncertainty of $Z_A$ normalized at infinite volume of
order 30\%. The uncertainty, however, has little effect in 
a Pad\'e fit of $Z_A$ and hence final results;    
$Z_A$ varies less than 3\% at the largest coupling $\beta=2.187$,
even if we artificially shift $Z_A$ at $\beta=2.4$ by 30\%.

\section{Results for $Z$-factors}\label{sec:results}

\begin{figure}[t]
\centerline{\epsfxsize=6.4cm \epsfbox{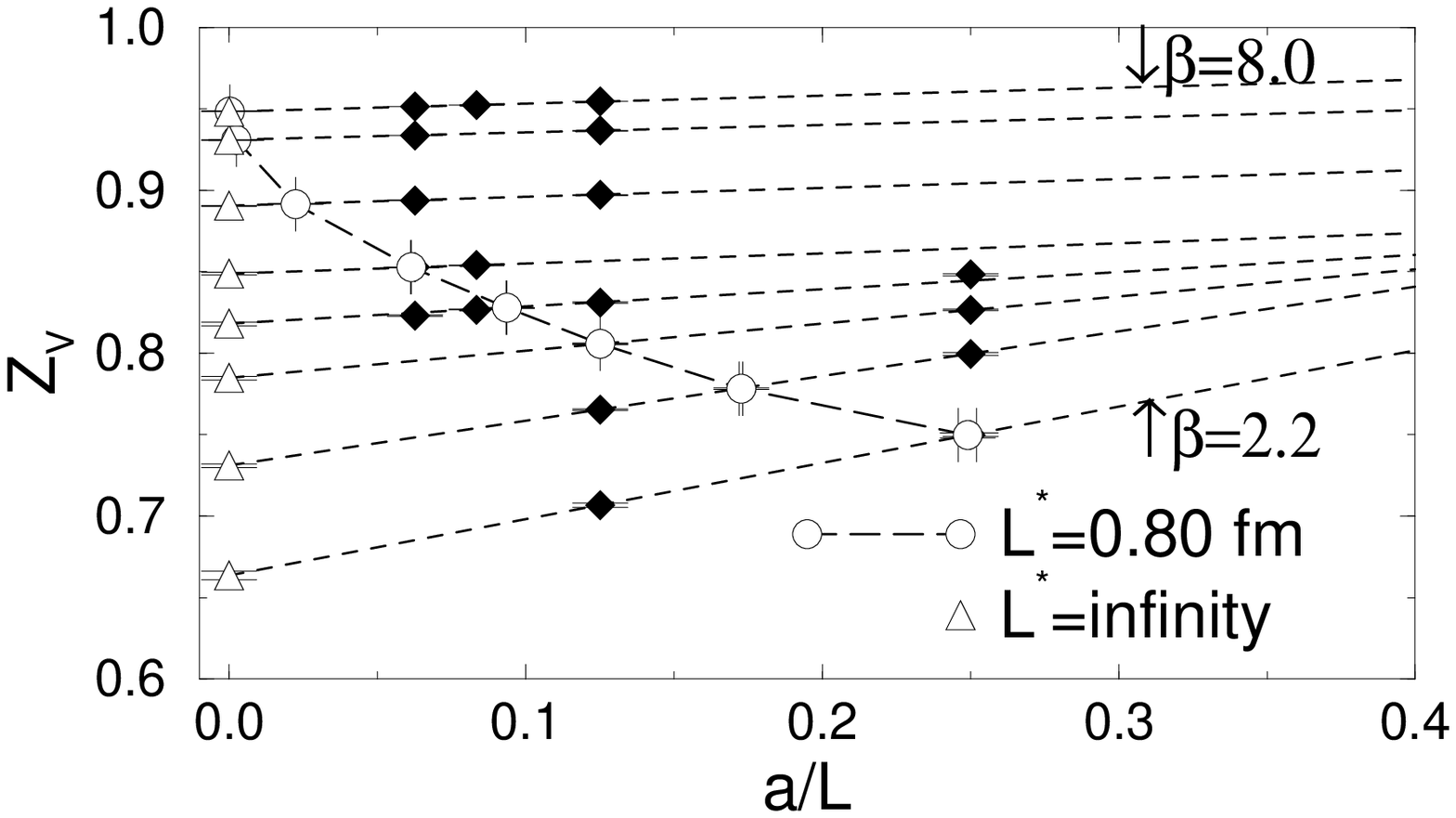}}
\centerline{\epsfxsize=6.4cm \epsfbox{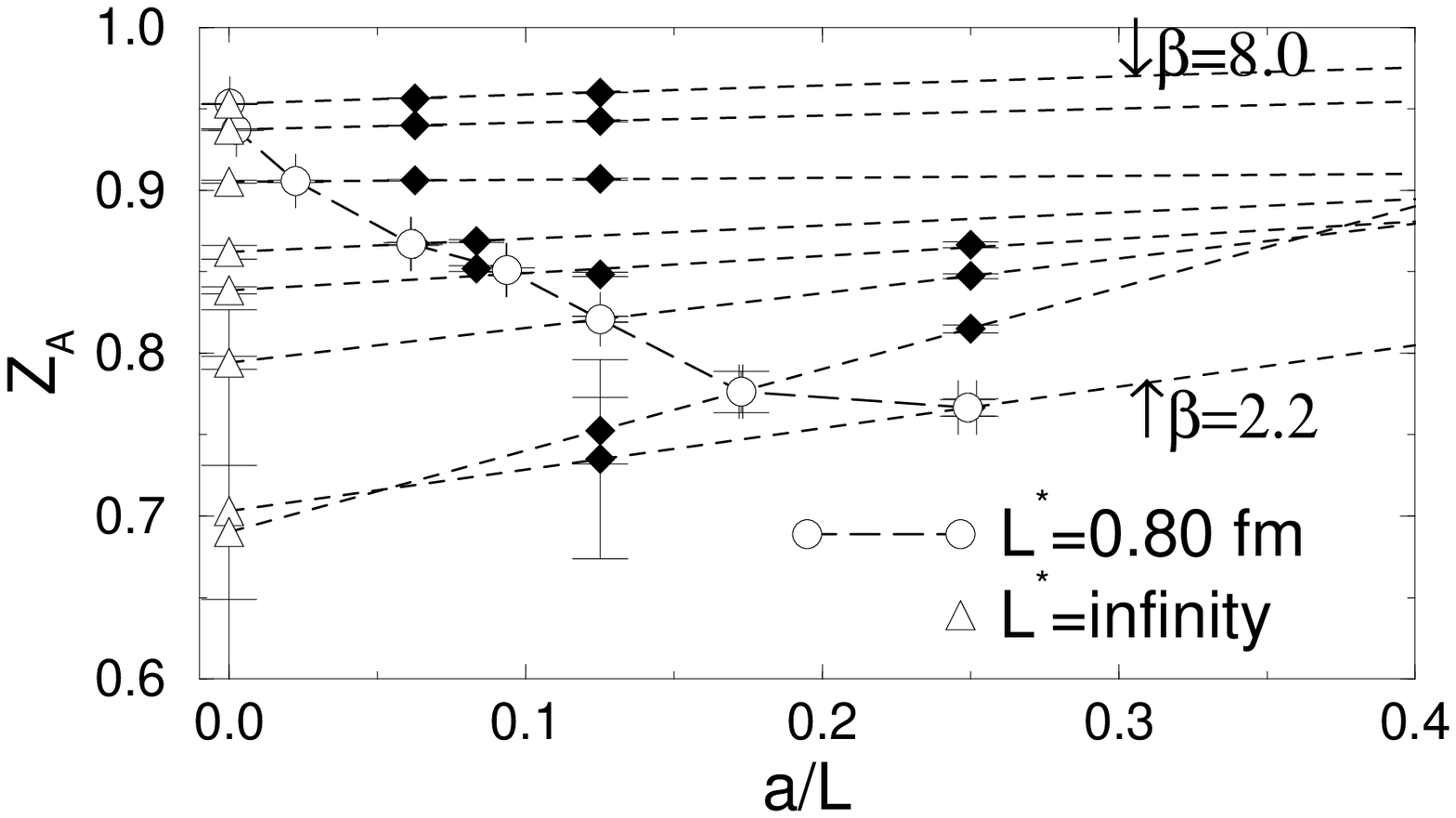}}
\vspace{-1cm}
\caption{Size dependence of $Z_V$ and $Z_A$.
}
\label{fig:Size-Dep-Z}
\vspace{-0.7cm}
\end{figure}

\begin{figure}[t]
\centerline{\epsfxsize=6.4cm \epsfbox{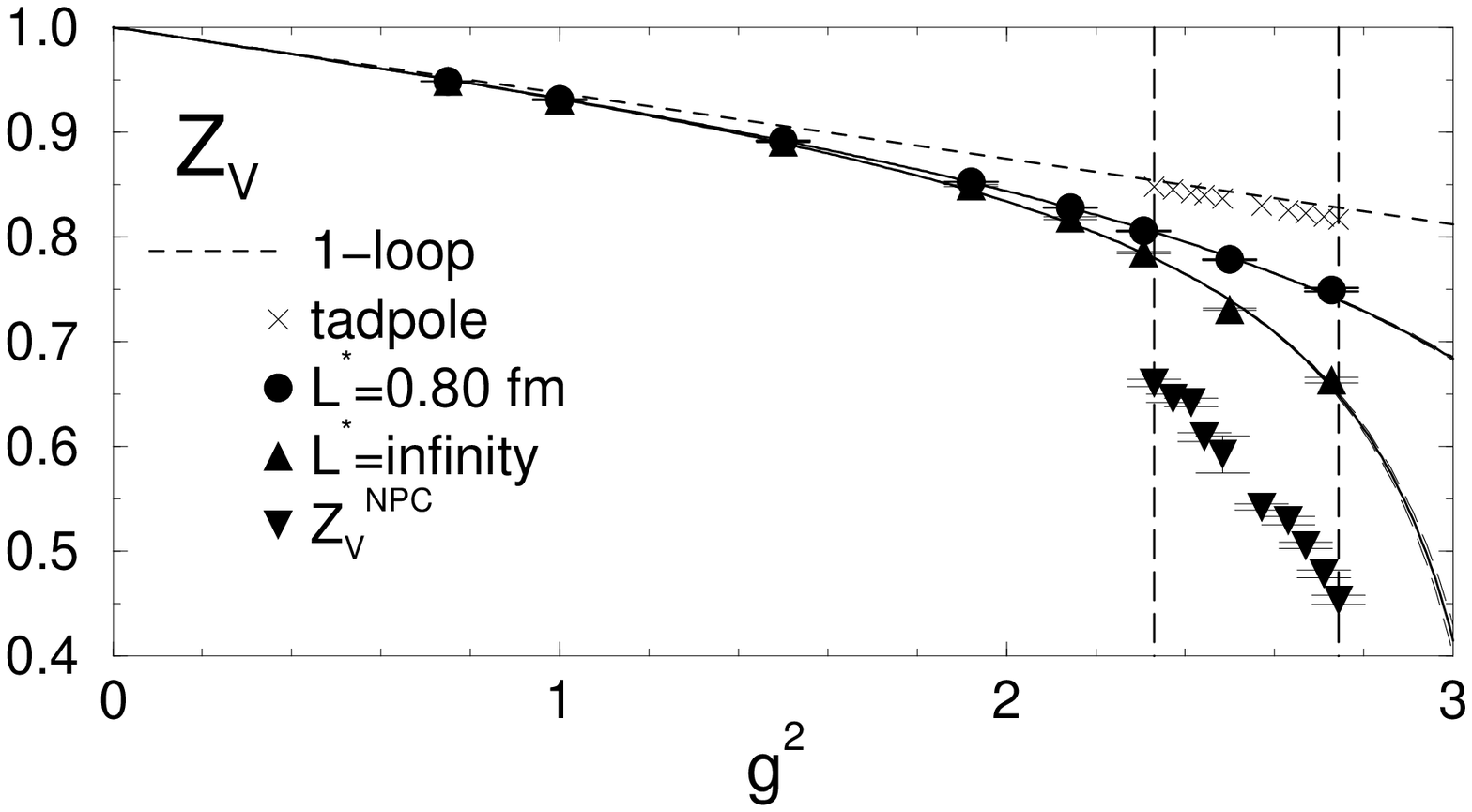}}
\centerline{\epsfxsize=6.4cm \epsfbox{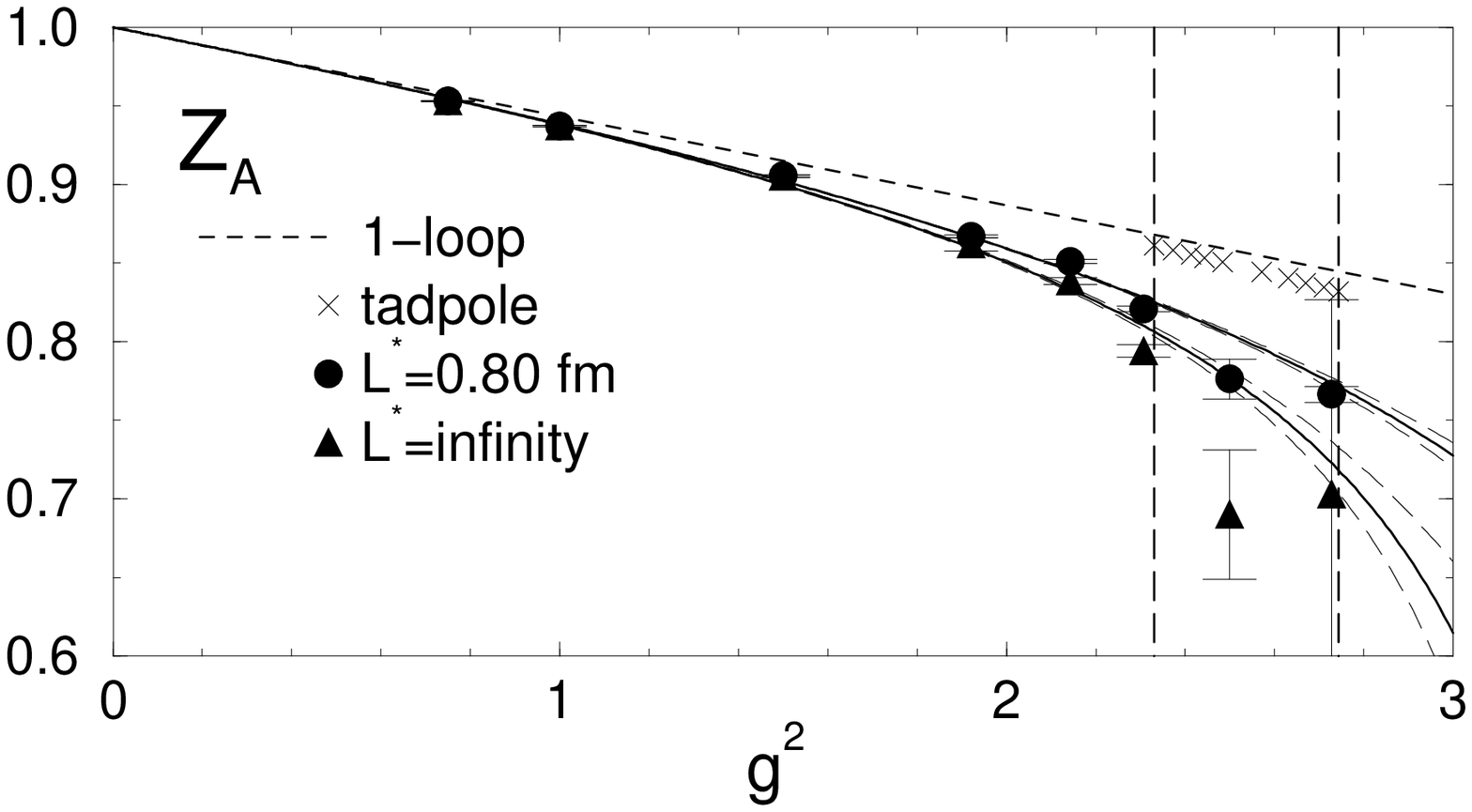}}
\vspace{-1cm}
\caption{Results for $Z_V$ and $Z_A$ normalized at 
$L^*=0.8$ fm and $L^*=\infty$. 
}
\label{fig:result}
\vspace{-0.7cm}
\end{figure}

We determine the $Z$-factors for infinite volume ($L^*=\infty$) 
and also for a fixed finite physical volume ($L^*=0.8$ fm corresponding to 
$8^3$ lattice at $\beta=$ 2.6) for comparison. 
The lattice scale is set through the string tension

As shown in Fig.~\ref{fig:Size-Dep-Z}, size dependence of $Z$-factors
becomes sizable toward strong couplings.
Since our quark action employs a tadpole-improved value of $c_{\rm sw}$, 
we expect $O(a)$ errors in the $Z$-factors. 
Therefore we extrapolate or interpolate results linearly in $a/L$ to 
obtain estimates at $L^*=0.8$~fm and at $L^*=\infty$.

In Fig.~\ref{fig:result} we show results of $Z$-factors as a function of 
bare coupling $g^2$, 
together with Pad\'e fits (solid curves in the figure) to them. 
Non-perturbative estimates give values smaller than the one-loop 
perturbative ones (dashed lines) by about 20 \% (15\%) for $Z_V$ ($Z_A$) 
at the largest coupling of the CP-PACS simulation, $\beta=2.187$.

$Z_V^{NPC}$ determined from the ratio of the conserved
vector current to the local one differs significantly from
$Z_V$ from the SF method, because the local current is not $O(a)$-improved.

\section{Scaling Property of Decay Constants}\label{sec:scaling}

\begin{figure}[t]
\centerline{\epsfxsize=6.4cm \epsfbox{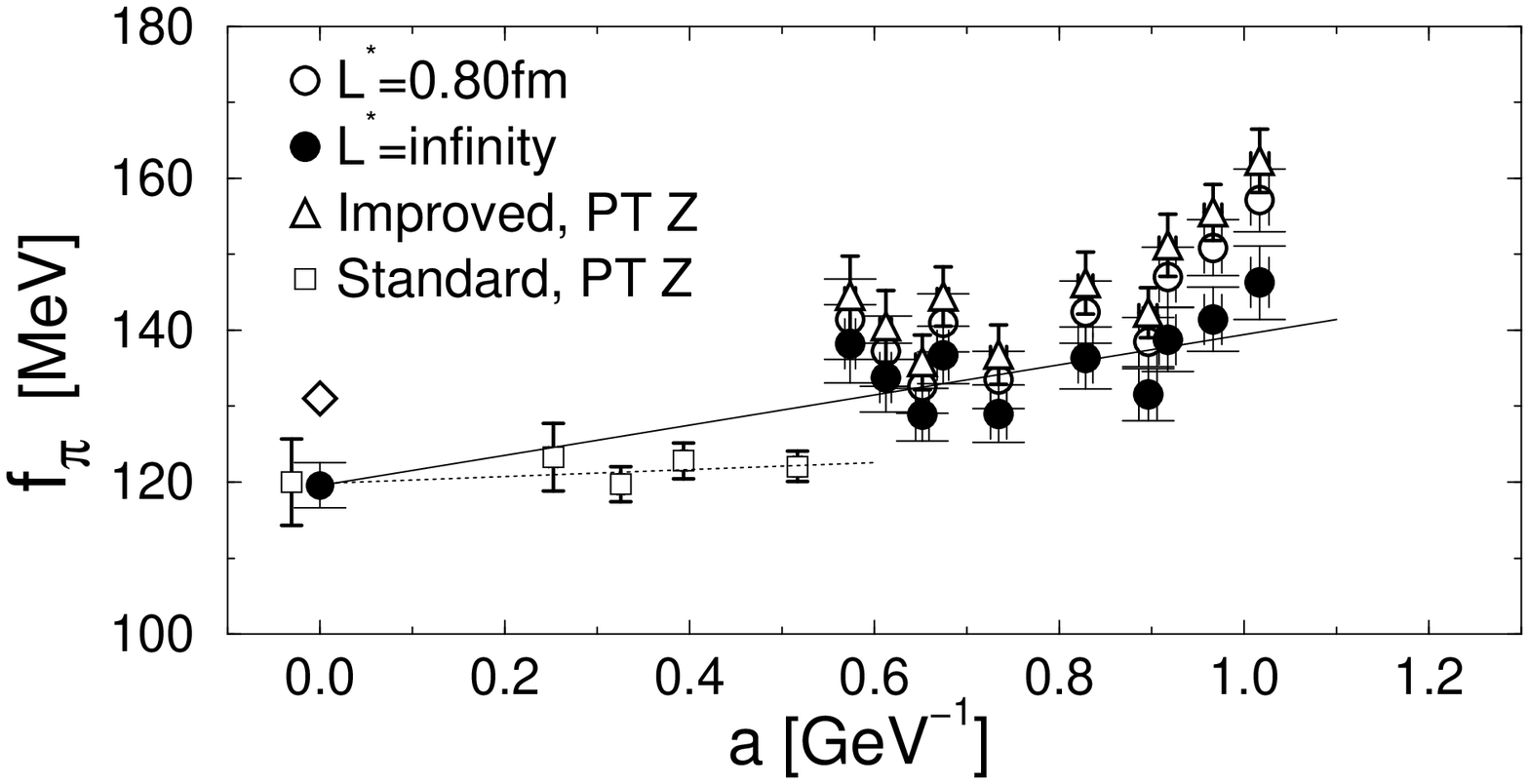}}
\centerline{\epsfxsize=6.4cm \epsfbox{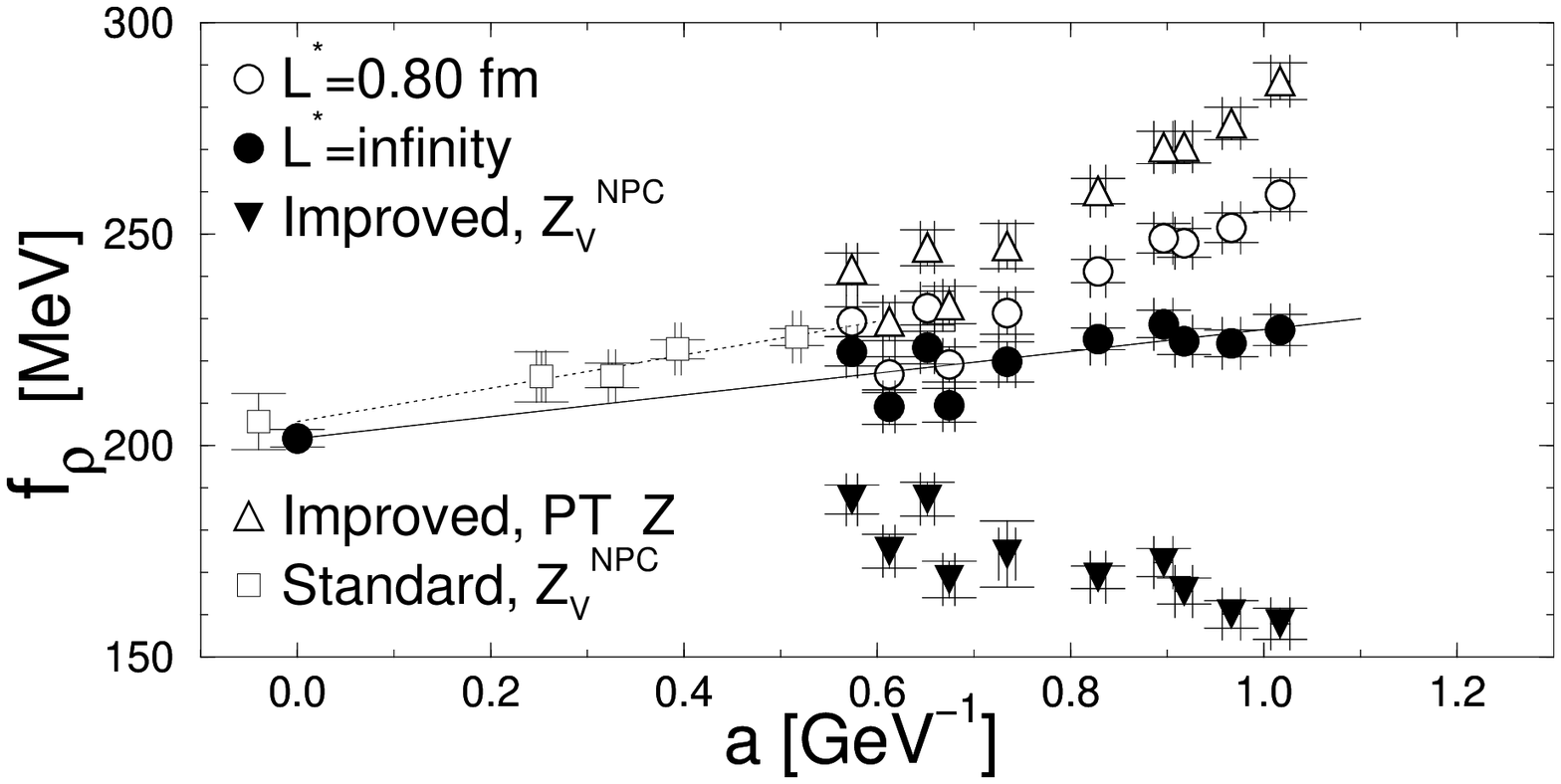}}
\vspace{-1cm}
\caption{$f_\pi$ and $f_\rho$ vs. $a$ for
our improved action with non-perturbative and perturbative
(PT) $Z$-factors together with results for the standard 
action~\protect\cite{ref:CPPACS-quench}.}  
\label{fig:decay}
\vspace{-0.8cm}
\end{figure}

We compare in Fig.~\ref{fig:decay} $f_\pi$ and $f_\rho$
determined with non-perturbative $Z$-factors normalized 
at $L^*=\infty$ (filled circles) with those using 
perturbative $Z$-factors (open up triangles).  
For comparison, open squares are 
the results from the standard plaquette and Wilson 
action\cite{ref:CPPACS-quench}.

We observe a very encouraging result that with the non-perturbative 
$Z$-factors scaling violations are sizably reduced.
Furthermore the continuum extrapolation yields values consistent with 
those from the standard action. 

In the same figure, we overlay $f_\pi$ and $f_\rho$ determined
with $Z$-factors normalized at finite $L^*=0.8$ fm 
(open circles). Scaling is best improved
when $Z$-factors are normalized at $L^*=\infty$. This property
is likely related to the fact that $O(a/L)$ errors in $Z$-factors
are removed in the limiting procedure $L^*\to \infty$.

We also find that $f_\rho$ determined from the conserved vector
current (filled down triangle in Fig.~\ref{fig:decay}) exhibits
a large scaling violation.


\vspace{2mm}
This work is supported in part by Grants-in-Aid of the Ministry of Education 
(Nos.
12304011, 
12640253, 
13135204, 
13640260, 
14046202, 
14740173, 
15204015, 
15540251, 
15540279  
).

\end{document}